\documentclass{article}
\usepackage{float}
\usepackage{hyperref}
\usepackage{algorithm}
\usepackage{algpseudocode}
\usepackage{arxiv}
\usepackage{amsmath}
\usepackage[utf8]{inputenc} % allow utf-8 input
\usepackage[T1]{fontenc}    % use 8-bit T1 fonts
\usepackage{hyperref}       % hyperlinks
\usepackage{url}            % simple URL typesetting
\usepackage{booktabs}       % professional-quality tables
\usepackage{amsfonts}       % blackboard math symbols
\usepackage{nicefrac}       % compact symbols for 1/2, etc.
\usepackage{microtype}      % microtypography
\usepackage{lipsum}
\usepackage{graphicx}
\graphicspath{ {./images/} }
\title{Generalized first order rational ODE reduction algorithm with bounded degree transformation}

\author{
 Shaoxuan Huang \\
  Chinese university of Hongkong, ShenZhen
  \texttt{223010044@link.cuhk.edu.cn} \\
}
\begin{document}
\maketitle
\begin{abstract}
The integrability problem of rational first-order ODEs $y^{\prime}=\frac{M(x,y)}{N(x,y)}$, where $M,N \in \mathbb{R}[x,y]$ is a long-term research focus in the area of dynamical systems, physics, etc. Although the computer algebra system such as Mathematica, Maple has developed standard algorithms to tackle its first integral expressed by Liouvillian or special function, this problem is quite difficult and the general method requires specifying a tight degree bound for the Darboux polynomial. Computing the bounded degree first integral, in general, is very expensive for a computer algebra system\cite{duarte2021efficient}\cite{cheze2020symbolic} and becomes impractical for ODE of large size. In \cite{huang2025algorithm}, we have proposed an algorithm to find the inverse of a local rational transformation $y \to \frac{A(x,y)}{B(x,y)}$ that transforms a rational ODE to a simpler and more tractable structure $y^{\prime}=\sum_{i=0}^nf_i(x)y^i$, whose integrability under linear transformation $\left\{x \to F(x),y \to P(x)y+Q(x)\right\}$ can be detected by Maple efficiently \cite{CHEBTERRAB2000204}\cite{cheb2000first}. In that paper we have also mentioned when $M(x,y),N(x,y)$ of the reducible structure are not coprime, canceling the common factors in $y$ will alter the structure which makes that algorithm fail. In this paper, we consider this issue. We conclude that with the exact tight degree bound for the polynomial $A(x,y)$ given, we have an efficient algorithm to compute such transformation and the reduced ODE for "quite a lot of" cases where $M,N$ are not coprime. We have also implemented this algorithm in Maple and the code is available in \href{https://www.researchgate.net/publication/393362858_Generalized_ODE_reduction_algorithm_docmw}{Researchgate}. 
\end{abstract}
\keywords{rational ODE, Liouvillian first integral, non-Liouvillian first integral, rational transformation, Symbolic computation}
\section{Review of the existing method}
The rational differential equation, which can be viewed as a planar autonomous system, takes the form:
\begin{equation}
    \frac{dy}{dx} = \frac{M(x,y)}{N(x,y)}
\end{equation}
and remains an important research interest in the study of dynamical systems and physics. For the methods for tackling the first integral of first order ODE, see \cite{goriely2001integrability, kamke2013differentialgleichungen} for an overview. \\
A special type of (1) attracting particular interest of researchers is:
\begin{equation}
    y^{\prime}=\sum_{i=0}^nf_i(x)y^i
\end{equation}
where $f_i(x)$ are rational functions in $x$. Containing Bernoulli, Chini, Riccati, Abel ODE as special cases, the integrability of such ODE can be characterized by equivalent classes under linear transformation, see \cite{appell1889invariants}\cite{CHEBTERRAB2000204}\cite{cheb2000first} for details: \\
\begin{equation}
    \left\{x \to F(x),y \to P(x)y+Q(x)\right\}
\end{equation}
In \cite{huang2025algorithm} we consider a more general problem, we consider a more general transformation applied to (2):
\begin{equation}
    \left\{x \to F(x), y \to \frac{A(x,y)}{B(x,y)}\right\}
\end{equation}
Applying such transformation yields an ODE of a more complicated structure:
\begin{equation}
    y^{\prime}=\frac{\sum_{i=0}^{n}f_iA^{i}B^{n-i}-B^{n-2}(B\frac{\partial A}{\partial x}-A\frac{\partial B}{\partial x})}{B^{n-2}(B\frac{\partial A}{\partial y}-A\frac{\partial B}{\partial y})}
\end{equation}
In \cite{huang2025algorithm} we assume that rational ODE (1) is exactly of form (5) and the numerator and denominator are coprime. We develop new approaches to detect the transformation $y \to \frac{A(x,y)}{B(x,y)}$ and if we find such transformation, we reduce the ODE to (2). However, if there exists an irreducible factor $p_i(x,y)$ of $A(x,y)$ with multiplicity $\alpha_i>1$ and $f_0(x) = 0$, then $p_i(x,y)^{\alpha_i-1}$ will be a common factor of the numerator and the denominator. After canceling the $p_i(x,y)^{\alpha_i-1}$, the ODE will no longer maintain structure (5) and the original algorithm fails in detecting the reducibility. \\
In this paper we are interested in the following problem: given $M,N$ in a rational ODE and the precise degree bound of $A(x,y)$, we aims to check whether there exist polynomials $A(x,y),B(x,y),f_i(x),t(x),c(x,y),n>2$, where $A,B$ are coprime bivariate polynomials in $\mathbb{R}[x,y]$, $c(x,y)|A(x,y)$, such that:
\begin{equation}
    \begin{aligned}
        & Mc = \sum_{i=0}^nf_iA^iB^{n-i}-t(B\frac{\partial A}{\partial x}-A \frac{\partial B}{\partial x})B^{n-2} \\
        & Nc = t(B\frac{\partial A}{\partial y}-A \frac{\partial B}{\partial y})B^{n-2} \\
    \end{aligned}
\end{equation}
Note that when $f_0(x) = 0$, $A(x,y)$ is a Darboux polynomial of the rational ODE $y^{\prime}=\frac{M(x,y)}{N(x,y)}$. Generally speaking, the degree upper bound of Darboux polynomials/first integrals is difficult to determine. A question of particular interest to researchers is: Given a relatively tight upper bound, does there exist an algorithm with acceptable time complexity/high efficiency to compute Darboux polynomials and first integrals? In general, determining the Darboux polynomials requires solving a large system of polynomial equations\cite{cheze2020symbolic}\cite{duarte2021efficient}. The computation cost of the existing Groebner bases/regular chains method for solving such polynomial equations grows very quickly as the size of the rational ODE increases. In this paper, we show that with assumption (6) for $M(x,y),N(x,y)$, we can compute $A(x,y),B(x,y)$ by solving linear equations and reduce the ODE to (2), which is much easier for determining the first integral/Darboux polynomial than the general form (1).
\section{New algorithm considering canceled factor}
\subsection{description of the new method}
The key idea of the new algorithm is to consider the canceled factor and $A$ separately, although in our assumption the canceled factor is also a factor of $A$. Since it is a factor of $A$, its degree in $y$ must be strictly less than the degree of $y$ in $A$. So we could search for the degree of canceled factor in $y$ from 0 to the degree of $A$ in $y$ minus 1. With each value in this range, we add the degree of cancelled factor in $y$ to the degree of $M,N$ in $y$, to recover the degree before common factors are canceled and $M,N$ are exactly of form (5). With these degrees, we can use the degree test procedure described in \cite{huang2025algorithm} to find the possible values of $n$ and candidates for $B$. For each possible candidate, consider solving the following equality for the undetermined coefficients in $c(x,y)$ and $A(x,y)$:
\begin{equation}
     Nc = t(B\frac{\partial A}{\partial y}-A \frac{\partial B}{\partial y})B^{n-2} 
\end{equation}
these coefficients satisfy a linear equation system, and for quite a lot of cases this is enough for determining the $A(x,y),B(x,y)$ as well as canceled factor $c(x,y)$(we will discuss counterexample in the next section). And with $A(x,y),B(x,y),c(x,y)$ we can continue to solve for $f_i(x)$ in the reduced ODE. Here we provide a short pseudo code of this algorithm:
\begin{scriptsize}
\begin{algorithm}[H]
    \caption{reduction algorithm for ODE $y^{\prime}=\frac{M(x,y)}{N(x,y)}$}
    \begin{algorithmic}[1]
        \Statex \textbf{Input:} Polynomial $M(x,y)$ and $N(x,y)$ in a rational ODE $y^{\prime}=\frac{M(x,y)}{N(x,y)}$ and the exact bound degreeA for A(x,y)
        \Statex \textbf{Output:} A reduced ODE $y^{\prime}t(x)=\sum_{i=0}^nf_i(x)y^i$ and the reduction transformation $y \to \frac{A(x,y)}{B(x,y)}$ when n>2
        \Statex i := degree(M,y)
        \Statex j := degree(N,y)
        \Statex t(x) := product of all factors of $N$ that depend on $x$ only
        \Statex factorList := all factors that depend on $y$
        \Statex multList := multiplicities of all factors that depend on $y$
        \Statex Construct candidates of A(x,y) with undetermined coefficients
        \Statex \textbf{for} cany \textbf{from} 0 \textbf{to} degreeA-1 \textbf{do:}
        \Statex \quad inew := i+cany;
        \Statex \quad jnew := j+cany;
        \Statex \quad Construct candidates of canceled factor of degree cany in y
        \Statex \quad \textbf{for} index1 \textbf{from} 3 \textbf{to} $\max(\text{multList})+2$ \textbf{do:}
        \Statex \quad \quad \textbf{for} index2 \textbf{from} $\max(\frac{\text{inew}}{\text{index1}},\frac{\text{jnew+1}}{\text{index1}})$ \textbf{to} jnew \textbf{do:}
        \Statex \quad \quad \quad \textbf{for} index 3 \textbf{from} 0 \textbf{to} index2-1 \textbf{do:}
        \Statex \quad \quad \quad \quad \textbf{if} index2+index3-1+(index1-2)index3=jnew \textbf{and} inew = index1 $\cdot$ index2 \textbf{and} index2 > cany \textbf{then:}
        \Statex \quad \quad \quad \quad \quad Choose possible factors of B
        \Statex \quad \quad \quad \quad \quad Construct candidates of B of degree index3
        \Statex \quad \quad \quad \quad \quad Verifying candidates of B and solve for A and canceled factor
        \Statex \quad \quad \quad \quad \quad use B,A,canceled factor to solve for $f_i(x)$
        \Statex \quad \quad \quad \quad \textbf{end if}
        \Statex \quad \quad \quad \quad \textbf{if} index2+index3-1+(index1-2)index2=jnew \textbf{and} (inew = index1 $\cdot$ index2 \textbf{or} inew $\leq$ jnew+1) \textbf{and} index3 $\geq$ cany \textbf{then:}
        \Statex \quad \quad \quad \quad \quad Choose possible factors of B
        \Statex \quad \quad \quad \quad \quad Construct candidates of B of degree index2
        \Statex \quad \quad \quad \quad \quad Verifying candidates of B and solve for A and canceled factor 
        \Statex \quad \quad \quad \quad \quad use B,A,canceled factor to solve for $f_i(x)$
        \Statex \quad \quad \quad \quad \textbf{end if}
        \Statex \quad \quad \quad \textbf{end do}
        \Statex \quad \quad \textbf{end do}
        \Statex \quad \textbf{end do}
        \Statex \textbf{end do}
    \end{algorithmic}
\end{algorithm}
\end{scriptsize}
\textbf{Example 2.1} Consider the following Abel differential equation, which is a representative of class 4 and solved in terms of Liouvillian function:
$$
    y^{\prime}=y^3-\frac{x+1}{x}y^2
$$
we apply the following transformation:
$$
     \left\{y \to \frac{(y+x+1)^2(y^2+x-1)}{(xy-2)^2(y+x^2-1)^2},x \to x \right\}
$$
which yields an ODE $y^{\prime}=\frac{M(x,y)}{N(x,y)}$ of degree 19, where:
$$
M= 4 x^{13} y^{6}+6 x^{12} y^{7}+ \cdots +7 x -4 y -4
$$
$$
N = 2 x \left(x^{12} y^{6}-x^{14} y^{3}+ \cdots +8 x +112 y -32\right)
$$
Assume polynomial A is of degree 4, use the algorithm we successfully obtain candidate for canceled factor and B at $n = 3$:
$$
c = x^{3} y b_{3,1}+x^{3} b_{3,0}+x^{2} y b_{2,1}+x^{2} b_{2,0}+x y b_{1,1}+x b_{1,0}+y b_{0,1}+b_{0,0}
$$
$$
B = (xy-2)^2(y+x^2-1)^2
$$
Assume $A = x^{4} a_{4,0}+x^{3} y a_{3,1}+x^{2} y^{2} a_{2,2}+x \,y^{3} a_{1,3}+y^{4} a_{0,4}+x^{3} a_{3,0}+x^{2} y a_{2,1}+x \,y^{2} a_{1,2}+y^{3} a_{0,3}+x^{2} a_{2,0}+x y a_{1,1}+y^{2} a_{0,2}+x a_{1,0}+y a_{0,1}+a_{0,0}$, substitute into the condition:
$$
    Nc = t(B\frac{\partial A}{\partial y}-A \frac{\partial B}{\partial y})B, t=x
$$
yields the following solution:
$$
A = x^{2} y^{2} b_{1,0}+2 x \,y^{3} b_{1,0}+y^{4} b_{1,0}+x^{3} b_{1,0}+2 x^{2} y b_{1,0}+3 x \,y^{2} b_{1,0}+2 y^{3} b_{1,0}+x^{2} b_{1,0}-x b_{1,0}-2 y b_{1,0}-b_{1,0}
$$
$$
c = x b_{1,0}+y b_{1,0}+b_{1,0}
$$
it is clear that $b_{1,0}$ could be any value other than 0. If we choose $b_{1,0}=1$, and substitute into the equality:
$$
Mc = \sum_{i=0}^nf_iA^iB^{n-i}-t(B\frac{\partial A}{\partial x}-A \frac{\partial B}{\partial x})B^{n-2}
$$
we successfully solve for the reduced ODE:
$$
y^{\prime}x=y^3x-(x+1)y^2
$$
\subsection{cases where such method fails}
Whether this method could succeed in finding canceled factor and A will depend on "how much" the canceled factor contributes to the total degree of A. For the above case, the canceled factor's degree is 1, and the total degree of A is 4. So, it does not contribute too much to the total degree of A. The coefficients of the solution of $c$ is linearly dependent. Simply by choosing parameters such that $c$ is non-zero yields a valid solution. However, if the canceled factor's degree dominates other factors in A, by solving (7), we will obtain "too general" result. That is, not all the non-zero parametric solutions for $c$ are valid. We will have to carefully choose these parameters and the valid choice is, in general, non-trivial. We will use an example to illustrate this.  \\
\textbf{Example 2.2} Consider the above Abel differential equation, if we change the transformation being applied to:
$$
\left\{y \to \frac{(y+x+1)^4}{(xy-2)^3(y+x^2-1)},x \to x \right\}
$$
which yields an ODE $y^{\prime}=\frac{M(x,y)}{N(x,y)}$ of degree 17, where:
$$
M = x^{11} y^{6}+5 x^{10} y^{7}+ \cdots -303 x -32 y -8
$$
$$
N = x \left(-3 x^{11} y^{5}+x^{10} y^{6}- \cdots +160 x -512 y +320\right)
$$
The factor being canceled is of degree 3. However, if we set:
$$
A = x^{4} a_{4,0}+x^{3} y a_{3,1}+\cdots+x a_{1,0}+y a_{0,1}+a_{0,0}
$$
$$
c = x^{3} y^{3} b_{3,3}+x^{3} y^{2} b_{3,2}+\cdots+x b_{1,0}+b_{0,1} y +b_{0,0}
$$
and substitute the correct candidate $B = (xy-2)^3(y+x^2-1)$ into the equality $ Nc = t(B\frac{\partial A}{\partial y}-A \frac{\partial B}{\partial y})B$. The solution for $A,c$ are:
$$
A = b_{3,0} x^{4}+\left(\frac{3 b_{2,1}}{2}-\frac{b_{3,0}}{2}\right) x^{3} y + \cdots +\left(\frac{b_{2,1}}{3}+3 b_{3,0}\right) y +\frac{4 b_{2,1}}{3}-3 b_{3,0}
$$
$$
c = b_{3,0} x^{3}+b_{2,1} x^{2} y +b_{2,1} x \,y^{2}+\frac{b_{2,1} y^{3}}{3}+\left(\frac{5 b_{2,1}}{6}+\frac{b_{3,0}}{2}\right) x^{2}+\left(\frac{3 b_{2,1}}{2}+\frac{3 b_{3,0}}{2}\right) x y +b_{2,1} y^{2}+\left(\frac{7 b_{2,1}}{6}-\frac{b_{3,0}}{2}\right) x +b_{2,1} y +\frac{b_{2,1}}{3}
$$
The parameters in these solutions still require further selection in order to make $c|A$ and solve the actual reduced ODE. \\
It is also important to note that we directly assume $t(x)$ contains all the factors depend on $x$ only in $N$, but this may not be true. When there is $a \in \mathbb{C}$ such that $A(a,y)=B(a,y)$, $B\frac{\partial A}{\partial y}-A \frac{\partial B}{\partial y}$ will produce $x-a$ as a factor. So we will need to try combination of different factors in $x$ in polynomial $N$ as candidates for $t(x)$. We will not discuss that issue for simplicity.
\section{Code implementation and Sample output}
We have implemented this algorithm in Maple and the code implementation is available at \href{https://www.researchgate.net/publication/393288749_Generalized_ODE_reduction_algorithm_docmw}{Researchgate}. The core function \texttt{degree-test} takes $M,N$ in a ODE and the specified degree bound for A as input and output A,B in the transformation and the canceled factor as well as the reduced ODE. \\
We design an example to demonstrates the capabilities of this algorithm.
\begin{figure}[H]
    \centering
    \includegraphics[width=0.8\linewidth]{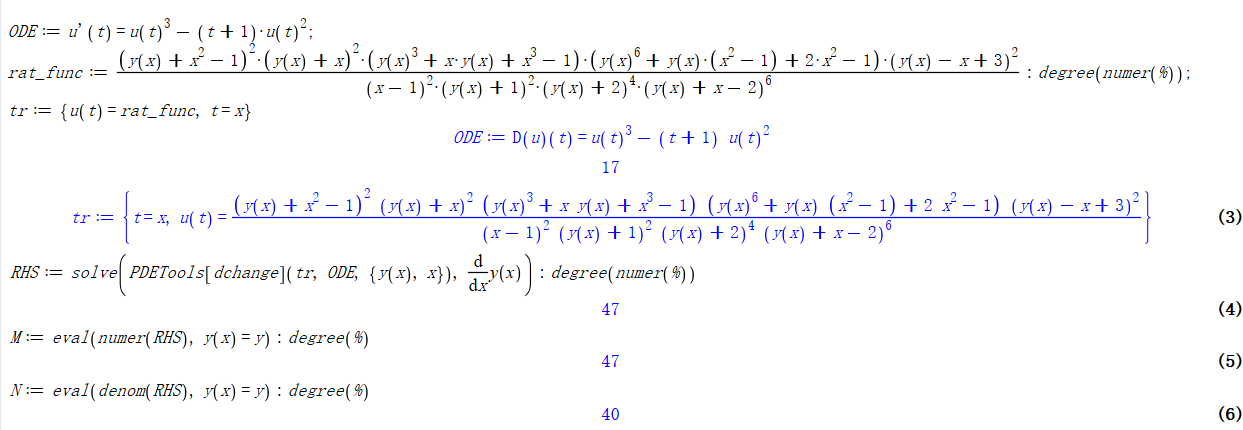}
    \caption{Contrive an ODE}
    \label{fig:enter-label}
\end{figure}
The rational ODE has degree 47 and the degree of A is 17. Executing \texttt{degree-test(M,N,17)} command, we derive an ODE of form (2) in 5.719 seconds. \\
\begin{figure}[H]
    \centering
    \includegraphics[width=0.8\linewidth]{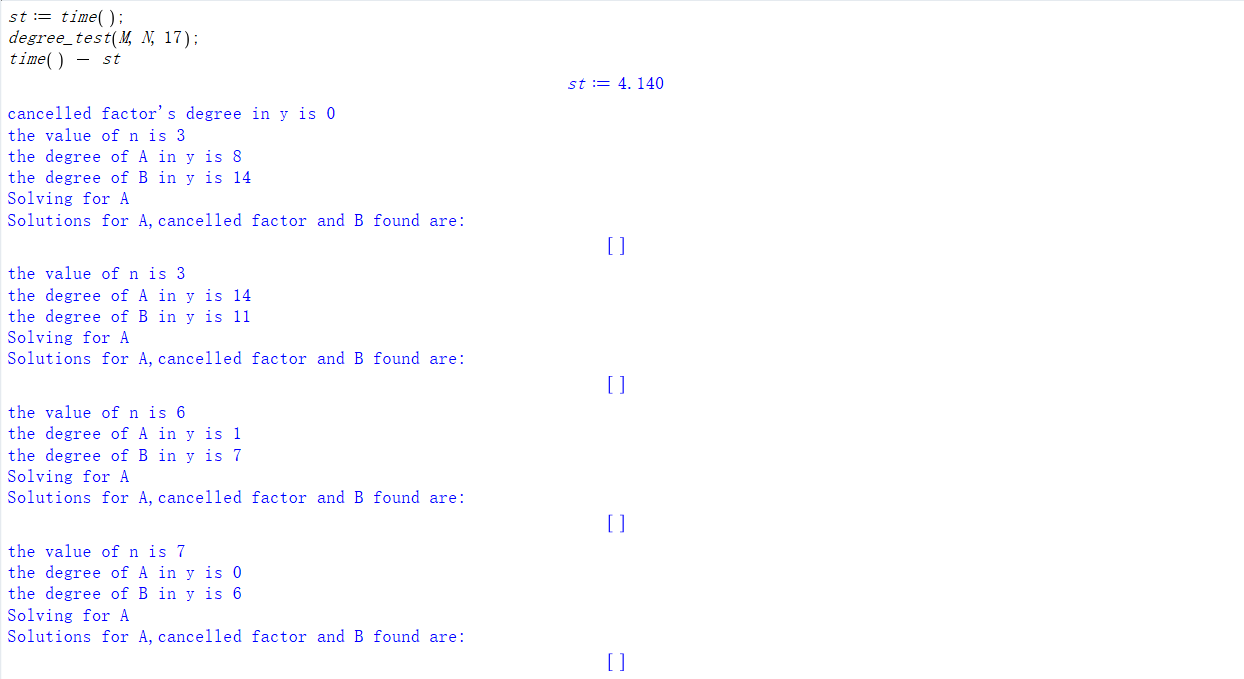}
    \caption{Excluding invalid degrees through candidate testing}
    \label{fig:enter-label}
\end{figure}
\begin{figure}[H]
    \centering
    \includegraphics[width=0.8\linewidth]{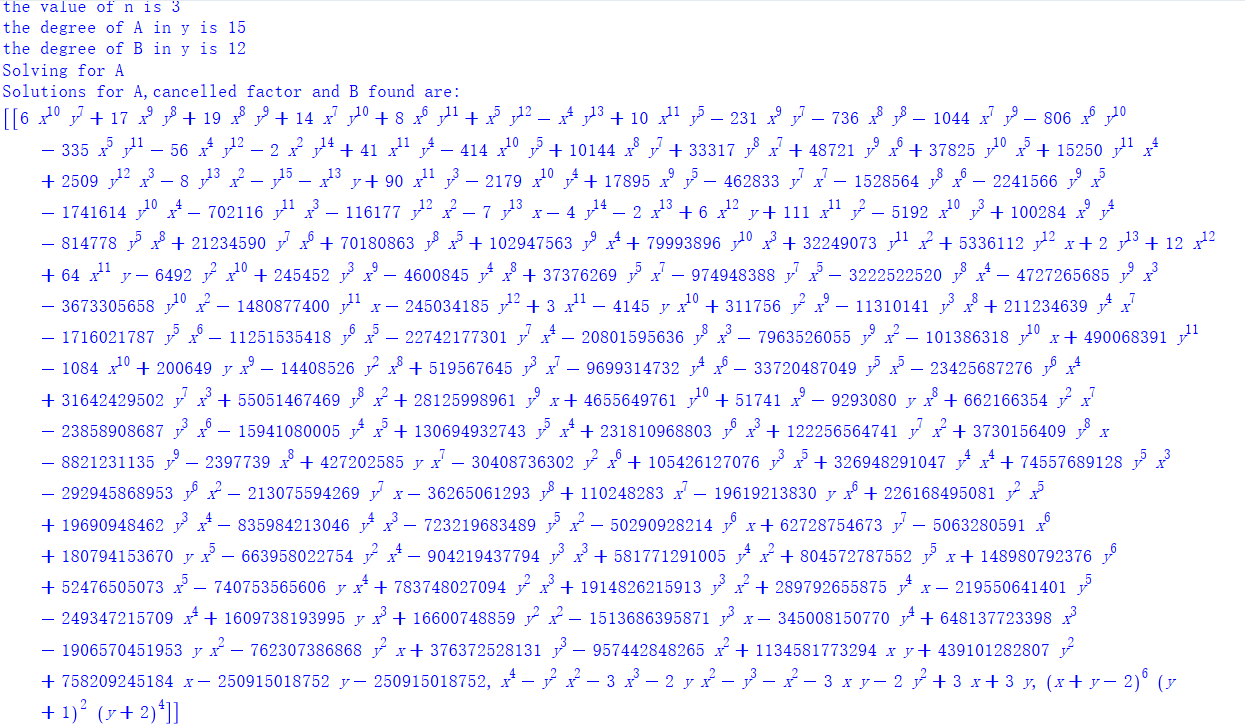}
    \caption{Finding A,canceled factor and B}
    \label{fig:enter-label}
\end{figure}
\begin{figure}[H]
    \centering
    \includegraphics[width=0.8\linewidth]{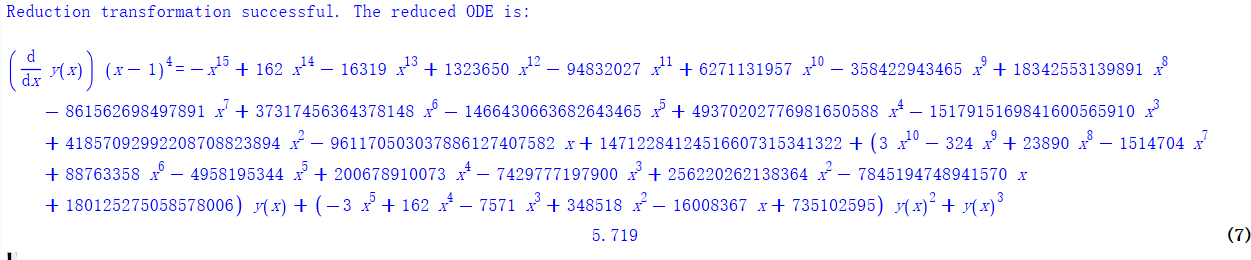}
    \caption{Successfully solve for reduced ODE}
    \label{fig:enter-label}
\end{figure}
Note that for ODE of form (2), it is sometimes easy to check for a Darboux polynomial of the form $y=F(x)$, where $F$ is a bounded degree polynomial. For the above example, we derive such Darboux polynomial using \texttt{PDETools[PolynomialSolutions]} in just 0.125 seconds. \\
\begin{figure}[H]
    \centering
    \includegraphics[width=0.8\linewidth]{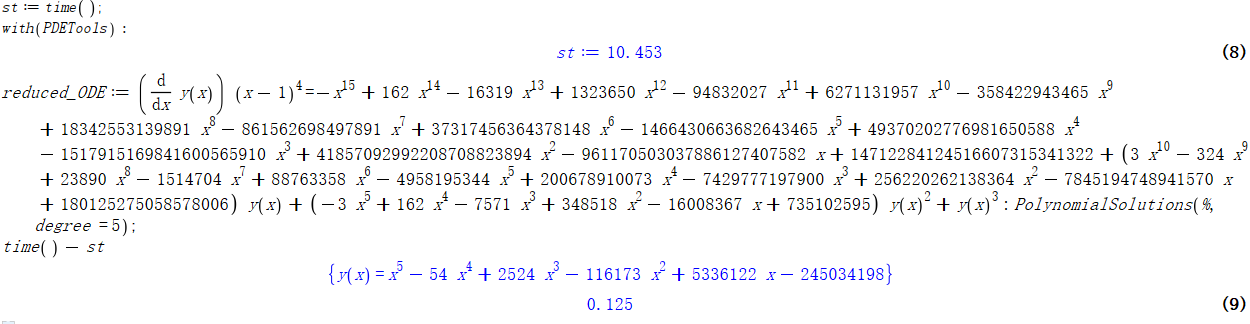}
    \caption{Finding the Darboux polynomial for ODE}
    \label{fig:enter-label}
\end{figure}
By design, the \texttt{ degree test(M,N,degreeA)} function only seeks a transformation of the input degree. If the user inputs a bound that is larger than the exact bound it will not return the result, since the parametric solution obtained from $Nc = t(B\frac{\partial A}{\partial y}-A \frac{\partial B}{\partial y})B$ contains "too general" result, as explained in section 2. However, the occurence of parametric solution is a good indication of the possibility of existence of lower degree transformation. To force the program print any parametric candidates found, set the forth argument \texttt{print-can} to be true, whose default value is false. Here is an example. \\
\begin{figure}[H]
    \centering
    \includegraphics[width=0.8\linewidth]{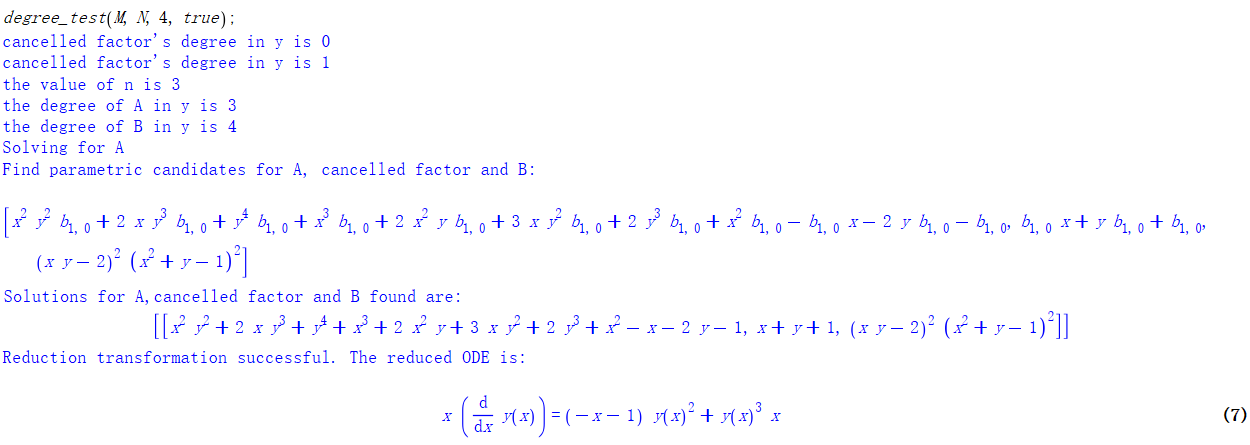}
    \caption{Print parametric candidates}
    \label{fig:enter-label}
\end{figure}
\bibliographystyle{plain}  
\bibliography{references}
\end{document}